\documentclass[12pt,preprint]{emulateapj}

\usepackage{graphicx,amsmath,natbib,color}

\usepackage{natbib}
\usepackage{color}
\usepackage{amsmath,amsthm,amssymb}
\usepackage{epsfig}
\usepackage{subfigure}
\usepackage{mathrsfs}
\usepackage{url}

\newcommand{\msun}{$M_{\odot}$}
\newcommand{\mjup}{$M_{Jup}$}
\newcommand{\av}{A$_{V}$}
\newcommand{\teff}{$T_{\rm eff}$}
\newcommand{\bcv}{BC$_{\rm V}$}

\slugcomment{Accepted to ApJ Letters on 13 May 2015}

\shorttitle{Discovery of Companions with Extreme Mass Ratios in Sco-Cen}
\shortauthors{Hinkley et al.}

\begin{document}

\title{Discovery of Seven Companions to Intermediate Mass Stars with Extreme Mass Ratios in the Scorpius-Centaurus Association\footnotemark[\lowercase{a}]}

\author{Sasha Hinkley\altaffilmark{1}}  
\author{Adam L. Kraus\altaffilmark{2}}
\author{Michael J. Ireland\altaffilmark{3}}
\author{Anthony Cheetham\altaffilmark{4}}
\author{John M. Carpenter\altaffilmark{5}}   
\author{Peter Tuthill\altaffilmark{4}}
\author{Sylvestre Lacour\altaffilmark{6}}
\author{Thomas M. Evans\altaffilmark{1}} 
\author{Xavier Haubois\altaffilmark{7,4}}
\altaffiltext{1}{University of Exeter, Physics Department, Stocker Road, Exeter, EX4 4QL, United Kingdom}
\altaffiltext{2}{Department of Astronomy, The University of Texas at Austin, Austin, TX 78712, USA}
\altaffiltext{3}{Research School of Astronomy \& Astrophysics, Australian National University, Canberra ACT 2611, Australia}
\altaffiltext{4}{Sydney Institute for Astronomy, School of Physics, The University of Sydney, NSW 2006, Australia}
\altaffiltext{5}{Department of Astronomy, California Institute of Technology, 1200 E. California Blvd., MC 249-17, Pasadena, CA 91125, USA }
\altaffiltext{6}{LESIA, CNRS/UMR-8109, Observatoire de Paris, UPMC, Universit\'e Paris Diderot, 5 place Jules Janssen, 92195 Meudon, France}
\altaffiltext{7}{European Southern Observatory (ESO), Alonso de Cordova 3107, Casilla 19001, Vitacura, Santiago 19, Chile}

\footnotetext[a]{Based on observations made with ESO Telescopes at the La Silla Paranal Observatory Under Program IDs:\,0.87.C-0790 and 089.C-0605}

\begin{abstract}
We report the detection of seven low mass companions to intermediate-mass stars (SpT B/A/F; $M$$\sim$1.5-4.5\,\msun) in the Scorpius-Centaurus Association using nonredundant aperture masking interferometry.  Our newly detected objects have contrasts $\Delta L'$$\approx$4--6, corresponding to masses as low as $\sim$20\,\mjup~and mass ratios of $q$$\sim$0.01-0.08, depending on the assumed age of the target stars. With projected separations $\rho$$\sim$10-30\,AU, our aperture masking detections sample an orbital region previously unprobed by conventional adaptive optics imaging of intermediate mass Scorpius-Centaurus stars covering much larger orbital radii ($\sim$30-3000\,AU). At such orbital separations, these objects resemble higher mass versions of the directly imaged planetary mass companions to the 10-30\,Myr, intermediate-mass stars HR\,8799, $\beta$ Pictoris, and HD\,95086.  These newly discovered companions span the brown dwarf desert, and their masses and orbital radii provide a new constraint on models of the formation of low-mass stellar and substellar companions to intermediate-mass stars. 
\end{abstract}

\keywords{instrumentation: adaptive optics---
instrumentation: spectrographs---
planets and satellites: detection---
techniques: high angular resolution
}

\section{Introduction}
Observing the population of planetary and brown dwarf companions orbiting young ($\sim$5-10 Myr) stars, soon after the dissipation of the primordial gaseous disk, is a key measure that will lend support to competing formation models of substellar objects \citep[e.g.][]{dcb04,sw09}.  Specifically, direct measurements of the {\it orbital} distribution of these objects shortly after formation \citep[e.g.][]{dvf09,kmy10} will serve as essential constraints to theoretical and numerical models of planetary formation. Thus, observing low mass companions as early as possible \citep[e.g.][]{kic14} will then serve as a ``snapshot'' of nascent system architecture, and largely eliminate any confusion about the initial conditions of companion formation caused by subsequent dynamical processes \citep[e.g.][]{sm09, cfr10}. Moreover, observing the luminosities of substellar companions in the first few million years is essential to constrain models of the intial entropy and temperatures of substellar objects  \citep[e.g.][]{fms08, mc14}.

\begin{deluxetable*}{cccccccccccc}
\tabletypesize{\scriptsize}
\tablecaption{Table of Observations}
\tablewidth{0pt}
\tablehead{ 
\colhead{Target} & 
\colhead{Region} & 
\colhead{P$_{\rm membership}$}  &   
\colhead{SpT} & 
\colhead{dist} & 
\colhead{$V$} &
\colhead{$W1$} & 
\colhead{\av} & 
\colhead{$M_{\rm bol}$} & 
\colhead{$\log$(\teff)} & 
\colhead{$M$} &
\colhead{Observatory}
\\
\colhead{(HIP)} & 
\colhead{ } & 
\colhead{(\%)} & 
\colhead{} & 
\colhead{(pc)} & 
\colhead{(mag)} & 
\colhead{(mag)} & 
\colhead{(mag)} & 
\colhead{(mag)} & 
\colhead{(K)} & 
\colhead{(\msun)} &  
\colhead{\& UT Date}
}
\startdata
71724 & UCL         &  94\%   & B8.5  & 157\tiny{$^{+16}_{-13}$}  & 6.63    & 6.82$\pm$0.05   & 0.04$\pm$0.23  & 0.03$\pm$0.55    & 4.072$\pm$0.087 & 3.41$^{+0.53}_{-0.75}$ &  VLT: 2011 June 10 \smallskip  \\ 
73990 & UCL         &  92\%   & A9     & 125\tiny{$^{+15}_{-12}$}  & 8.23    & 7.28$\pm$0.03   & 0.30$\pm$0.17  & 2.48$\pm$0.28    & 3.872$\pm$0.022 & 1.72$^{+0.08}_{-0.11}$ &  VLT: 2011 June 10 \smallskip  \\
74865 & UCL         &  96\%   & F4     & 115\tiny{$^{+19}_{-14}$}  & 9.00   & 7.78$\pm$0.03   & 0.20$\pm$0.07  & 3.50$\pm$0.31      & 3.822$\pm$0.006 & 1.42$^{+0.04}_{-0.06}$& VLT: 2011 June 10 \smallskip  \\
78196 & USco       &  95\%   & A0     & 127\tiny{$^{+9}_{-8}$}       & 7.03   & 7.09$\pm$0.05  & -0.05$\pm$0.25 & 1.36$\pm$0.45    & 3.987$\pm$0.076 & 2.46$^{+0.31}_{-0.59}$ &  Keck: 2010 May 24 \smallskip  \\
78233 & USco       &  79\%   & F0     &  145\tiny{$^{+15}_{-15}$}  & 9.17   & 7.64$\pm$0.03   & 0.84$\pm$0.11  & 2.55$\pm$0.25    & 3.857$\pm$0.012 & 1.67$^{+0.07}_{-0.12}$ & Keck:  2010 April 25 \smallskip  \\
79124 & USco       &  93\%   & A0     &  123\tiny{$^{+11}_{-9}$}    & 7.83   & 6.96$\pm$0.05  & 0.88$\pm$0.25  & 1.30$\pm$0.46      & 3.987$\pm$0.076 & 2.48$^{+0.32}_{-0.58}$ &  Keck:  2010 April 05 
\enddata
\tablecomments{``UCL'' and ``USco'' refer to the Upper Centaurus Lupus, and Upper Scorpius regions of Sco-Cen, respectively. Membership probabilies in column three are taken from \citet{rir11}}
\label{targtable}
\end{deluxetable*}

However, due to the scarcity of young stars in the solar neighborhood, assembling a statistically robust sample of low-mass companions requires observations of large, newly-formed stellar associations.  One such region, the Scorpius-Centaurus (hereafter ``Sco-Cen'') association, with a distance of $\sim$120-150\,pc \citep{dhd99} and 5-20\,Myr age \citep[e.g.~][]{pmb12,szb12} is the nearest OB Association. The young age of this association ensures that any planetary and brown dwarf companions will have elevated luminosity \citep[e.g.][]{bcb03}, allowing access to planetary mass objects even with observations achieving modest relative contrasts.

In addition to selecting young targets, observing stars more massive than solar (``intermediate mass stars'', 2-5\,\msun) may also enhance the probability of detection of substellar objects. Indeed, some studies suggest the fraction of super-Jupiter companions may be higher for stars more massive than $\sim$2\,\msun~\cite[][]{cj11, vpm12, rameau95086, rbq15}, possibly due to initially more massive circumstellar disks \citep[e.g.][]{ark13}, or possibly to serve as a reservoir for the conserved initial angular momentum of the star forming cloud \citep[e.g.][]{kbp07}.
The relatively short main-sequence lifetimes of these intermediate mass stars implies young associations such as Sco-Cen should have a greater fraction of these stars compared to the local solar neighborhood, making Sco-Cen a particularly promising region to study young, intermediate mass stars. 

To date, the only companions to Sco-Cen {\it intermediate-mass} stars lie at wide separations \citep[$\sim$30-3200\,AU, e.g.][]{kbp07, akl13, jlj13, ljv14,bmr14}.  The  $\sim$120--150\,pc distance to Sco-Cen stars means that orbital separations of substellar companions to these stars located near the ice line (5-10\,AU), where planet formation is thought to be most efficient \citep[e.g.][]{phb96}, corresponds to angular separations very close to the near-infrared diffraction limit of 10\,m telescopes ($\lambda/D\approx$ 30-45 miliarcsec).  Thus, sensitivity near, and within, the diffraction limit of large telescopes is needed to access smaller orbital separations for Sco-Cen stars \citep[][]{kim08, kim11}.

Aperture masking interferometry \citep[e.g.][and references therein]{tmd00,i13}, provides sensitivity at scales up to, and somewhat within, the usually defined diffraction limit ( $\sim$$\frac{1}{3}\lambda/D - 4\lambda/D \simeq$ 20-300 mas for Keck $L'$-band imaging). Applications of this technique \citep[e.g.,][and references therein]{ik08, hci11,ki12} use AO along with an opaque mask containing several holes, constructed such that the baseline between any two holes samples a unique spatial frequency in the pupil plane. Further, no coronagraphic mask is used, which avoids problems associated with measuring the relative astrometry between the occulted host star and a detected companion \citep[e.g.][]{dho06}.  Despite its very good sensitivity to small inner working angles, aperture masking interferometry usually only achieves typical contrasts of $\sim$5-8\,mag \citep{kim08,hci11}.  Nonetheless, these modest contrasts are still sensitive to planetary mass companions at young ages \citep{ik08,ki12}.

This paper presents discoveries of companions with extreme mass ratios, $q$=$M_{\rm companion}/M_{\rm host}$$\approx$0.01-0.08, from an ongoing multiplicity survey of $\sim$140 intermediate mass stars (SpT=B0-F2) in the Sco-Cen region using aperture masking interferometry.  We have selected our targets based on the refined Bayesian Sco-Cen membership selection technique described in \citet{rir11}, which uses radial velocity information to confirm or reject candidates.  Since high-mass stars frequently host one or more binary companions \citep[e.g.][]{dk13} which prevents the required contrast from being achieved, the targets in this study have been screened using past literature to eliminate binary systems with typical separations of $>$$30$\,mas and masses $>$$0.1$\,$M_\odot$.  Following this, we used results from our own ongoing snapshot AO imaging programs at Keck, VLT, and Palomar Observatories to eliminate other systems with evidence for binarity.  A more comprehensive discussion of our target selection and broad survey results will be given in a subsequent work (Rizzuto et al., {\it in prep}).  In \S\ref{starprops}, we describe the host star properties for our discoveries, followed by our observations and analysis strategy (\S\ref{obs}).  In \S\ref{results} and \S\ref{conclusions} we summarize our findings and place them in context. 

\begin{deluxetable*}{lrrrrrcccc}
\tabletypesize{\scriptsize}
\tablecaption{Properties of Newly Detected Companions }
\tablewidth{0pt}
\tablehead{ 
\colhead{Companion}  &   
\colhead{$\Delta L'$} &  
\colhead{Sep } & 
\colhead{PA} & 
\colhead{Confidence} &
\colhead{$M_{L'}$ } &
\colhead{10 Myr Mass } &  
\colhead{20 Myr Mass } &  
\colhead{Sep } &
\colhead{$q$=$M_{\rm comp}/M_{\rm star}$}  
\\
\colhead{} &
\colhead{(mag)} &
\colhead{(mas)} &
\colhead{(degrees)} &
\colhead{Level} &
\colhead{(mag)} & 
\colhead{(\mjup)} &
\colhead{(\mjup)} &
\colhead{(AU)} &
\colhead{(10\,Myr, 20\,Myr)}
}
\startdata
HIP71724B & 4.86\tiny{$^{+0.09}_{-0.08}$} & 103\tiny{$^{+4}_{-5}$}     & 202\tiny{$^{+2}_{-2}$}  & $>$99.9\%         &  5.70\tiny{$^{+0.33}_{-0.34}$} & 146{\tiny $^{+45}_{-32}$} & 217{\tiny $^{+67}_{-47}$} & 16\tiny{$^{+2}_{-2}$}  &  0.04{\tiny$_{-0.01}^{+0.03}$}, 0.06{\tiny$_{-0.02}^{+0.04}$}    \smallskip  \\
HIP73990B & 6.04\tiny{$^{+0.52}_{-0.34}$} & 161\tiny{$^{+12}_{-13}$} & 281\tiny{$^{+7}_{-6}$}  & $>$99.9\%         &  7.84\tiny{$^{+0.77}_{-0.62}$} & 21{\tiny $^{+30}_{-5 }$}    & 50{\tiny $^{+31}_{-33}$}   & 20\tiny{$^{+4}_{-3}$}  &  0.01{\tiny$_{-0.05}^{+0.02}$}, 0.03{\tiny$_{-0.02}^{+0.02}$}    \smallskip  \\  
HIP73990C & 5.95\tiny{$^{+0.55}_{-0.44}$} & 254\tiny{$^{+12}_{-38}$} &  48\tiny{$^{+13}_{-3}$} & $>$99.9\%         &  7.75\tiny{$^{+0.80}_{-0.72}$} & 22{\tiny $^{+38}_{-6 }$}    & 54{\tiny $^{+40}_{-30}$}   & 32\tiny{$^{+5}_{-7}$}  &  0.01{\tiny$_{-0.01}^{+0.02}$}, 0.03{\tiny$_{-0.02}^{+0.03}$}   \smallskip  \\  
HIP74865B & 5.03\tiny{$^{+0.28}_{-0.23}$} & 201\tiny{$^{+19}_{-12}$} & 115\tiny{$^{+3}_{-3}$}  & $>$99.9\%         &  7.51\tiny{$^{+0.59}_{-0.60}$} & 28{\tiny $^{+37}_{-10}$}   & 65{\tiny $^{+36}_{-25}$}   & 23\tiny{$^{+6}_{-4}$}  &  0.02{\tiny$_{-0.01}^{+0.03}$}, 0.04{\tiny$_{-0.02}^{+0.03}$} \smallskip  \\   
HIP78196B & 4.61\tiny{$^{+0.18}_{-0.22}$} & 74\tiny{$^{+10}_{-10}$}   & 266\tiny{$^{+4}_{-4}$}  & $\lesssim$99\% &  6.18\tiny{$^{+0.37}_{-0.42}$} & 98{\tiny $^{+42}_{-12}$}   & 152{\tiny $^{+56}_{-28}$} & 9\tiny{$^{+2}_{-2}$}     &  0.04{\tiny$_{-0.01}^{+0.03}$}, 0.06{\tiny$_{-0.02}^{+0.05}$} \smallskip  \\ 
HIP78233B & 4.72\tiny{$^{+0.11}_{-0.12}$} & 133\tiny{$^{+3}_{-3}$}     &  20\tiny{$^{+1}_{-1}$}   & 99.9\%               &  6.55\tiny{$^{+0.38}_{-0.36}$} & 86{\tiny $^{+12}_{-21}$}   & 124{\tiny $^{+27}_{-24}$}  & 19\tiny{$^{+2}_{-2}$} &  0.05{\tiny$_{-0.01}^{+0.01}$}, 0.07{\tiny$_{-0.02}^{+0.02}$} \smallskip  \\
HIP79124B & 4.30\tiny{$^{+0.10}_{-0.10}$} & 177\tiny{$^{+3}_{-3}$}     & 242\tiny{$^{+1}_{-1}$}  & 99.9\%               &  5.81\tiny{$^{+0.32}_{-0.34}$} &135{\tiny $^{+38}_{-33}$}  & 201{\tiny $^{+56}_{-44}$}  & 22\tiny{$^{+2}_{-2}$} &  0.05{\tiny$_{-0.02}^{+0.04}$}, 0.08{\tiny$_{-0.02}^{+0.05}$}   
\enddata
\tablecomments{Masses are derived from \citet{cba00} and \citet{bca98, bcb03} models.}
\label{contrasttable}
\end{deluxetable*}

\section{Target Star Properties}\label{starprops}
In Table\,\ref{targtable}, we list the basic properties of the targets described in this work.  With an ultimate goal of calculating host star mass, we start by calculating the bolometric magnitude ($M_{\rm bol}$) from the $V$ magnitude plus an estimate of the visual extinction \av~and bolometric correction \bcv. We first estimated the visual extinction for each of our targets by comparing the observed $(V-K)$ colors of our target stars with the $(V-K)$ colors for the corresponding spectral types tabulated by \citet{pmb12}.  As noted in Table\,\ref{targtable}, the spectral types we assume come either from the HD catalog \citep{h78} or from \citet{pm13}. Our spectral types have uncertainties of $\pm$2 subclass for the spectral types listed in the HD catalog, and $\pm$1 for those listed in \citet{pmb12}.  We assume that our uncertainties in spectral type result in $(V-K)$ color uncertainties of 0.1-0.2 mag, while the uncertainties in the color versus spectral type relation do not exceed 0.05 mag.  Next, using this estimated and the observed $(V-K)$ color, we estimate \av~using the relation between \av~and $A_K$ from \citet{sfd98}.  Using the spectral types and uncertainties, we adopt \bcv~values from \citet{pm13}, where the uncertainty in \bcv~is set by the uncertainty in the spectral type. We then combine $V$ mag, \av, \bcv, and the distance modulus to calculate bolometric absolute magnitudes $M_{\rm bol}$, and its uncertainty. Next, we estimate \teff~and its uncertainty using the SpT along with the tabulated values in \citet{pm13}. Lastly, we use the \teff~and $M_{\rm bol}$ values calculated above to calculate the mass for each target star by calculating a two-dimensional surface, mass(\teff, $M_{\rm bol}$), and then marginalizing the probability distribution function of (\teff, $M_{\rm bol}$) to obtain 1$\sigma$ confidence intervals. These masses are listed in Table\,\ref{targtable}.

As Table\,\ref{targtable} shows, all of the stars considered in this paper are members of either the Upper Centaurus Lupus (hereafter ``UCL'') or the Upper Scorpius (hereafter ``U Sco'') subgroups.  Rather than assign distances to each star that reflect the average distances of the Sco-Cen subgroups (e.g.\,125$\pm$15 pc and 145$\pm$15 pc for UCL and U Sco, respectively), we use the individual Hipparcos parallaxes recorded for these stars, since the individual parallax uncertainties associated with each target are comparable to the subgroup distance uncertainties ($\sim$10\%). However, HIP 78233 has a recorded Hipparcos parallax inconsistent with the median U Sco members, with large uncertainty (4.84$\pm$1.37\,mas). So we assign to it a distance of 145$\pm$15 pc, the median distance to U Sco.

While we have selected our targets partly based on their high probabilities of Sco-Cen membership as stated in \citet{rir11}, placement of the host stars on an HR diagram using the values calculated in Table\,\ref{targtable} verifies all targets can be well fit between the 10 and 20 Myr isochrones from \citet{bmg12}. These ages are broadly consistent with the reported ages of UCL and U Sco.  Some disagreement persists over the age of U Sco, which ranges from 5 Myr \citep{pz99,pbb02} to as high as 11 Myr \citep{pmb12}. Indeed, a 5\,Myr age is required to place low mass U Sco stars on an HR diagram \citep[Rizzuto et al. {\it in prep}, ][]{pm08, kcc15}. Further, when placed on an HR diagram at least half of our targets (HIP71724, HIP78196 and HIP79124) have positions in the HR diagram that are consistent with a 5\,Myr isochrone. Nonetheless, for overall consistency we report our derived masses using a 10-20 Myr age range.

\section{Observation Strategy \& Analysis}\label{obs}
All the data presented in this work were obtained at $L'$-band wavelengths (3.76 $\mu$m) using the NIRC2 infrared camera and AO system at the W. M. Keck Observatory using , as well as the ESO Very Large Telescope (``VLT'') NACO AO system and infrared camera. An observing sequence consisted of observing a target star in two opposed quadrants of the infrared camera: NIRC2 at Keck, and CONICA at the VLT.  A nine hole aperture mask  is used at Keck, and a seven hole mask was used at VLT, producing interferograms like that shown in Figure 1 of \citet{hci11}. At both detector positions, we typically obtained 15 images (30 images total) with an effective exposure time of 20s each. Usually two to four such 30-image sequences of each target star were obtained. 

We did not explicitly observe calibrator stars for each of our target stars, as is common practice.  Rather, we use all of the stars in a given observing night as mutual calibrators. Those with closure phase signals indicative of a companion are weighted lower in the list of calibrators than those without. In a single observing night, this method allows roughly twice the number of Sco-Cen targets to be observed.  To use the closure phase quantity to search for  companions, we follow the analysis outlined in \citet{kim08}, \citet{ik08} and \citet{hci11}, briefly summarized here. The data are initially flatfielded, sky subtracted, aligned, and corrected for cosmic rays.   The bispectrum, the complex triple product of visibilities defined by the three baselines formed from any three subapertures, is then calculated. The phase of this complex quantity is the closure phase.

As discussed in \citet{kim08} and \citet{hci11}, the calibrated object closure-phase is found by subtracting a weighted average of the closure-phase for the calibrator stars.  For the analysis in this paper, which is motivated by the search for point sources, the squared visibilities were not used as they were noisier than the closure-phases. Each target was calibrated against all other stars to search for any deviations from single point-like sources. The Root Mean Square (RMS) calibrated closure-phase was found for each of these target-calibrator pairs, and all calibrations that resulted in an RMS closure phase more than 1.5 times the minimum for each target over all calibrators were assigned a weight of zero. In practice, this meant that each target data set was calibrated by an average of the two or three calibrator data sets obtained closest in time. The lack of perfect closure phase calibration is still the dominant source of closure-phase noise in this analysis.

\begin{figure*}
  \centering
  \includegraphics[width=1.0\textwidth]{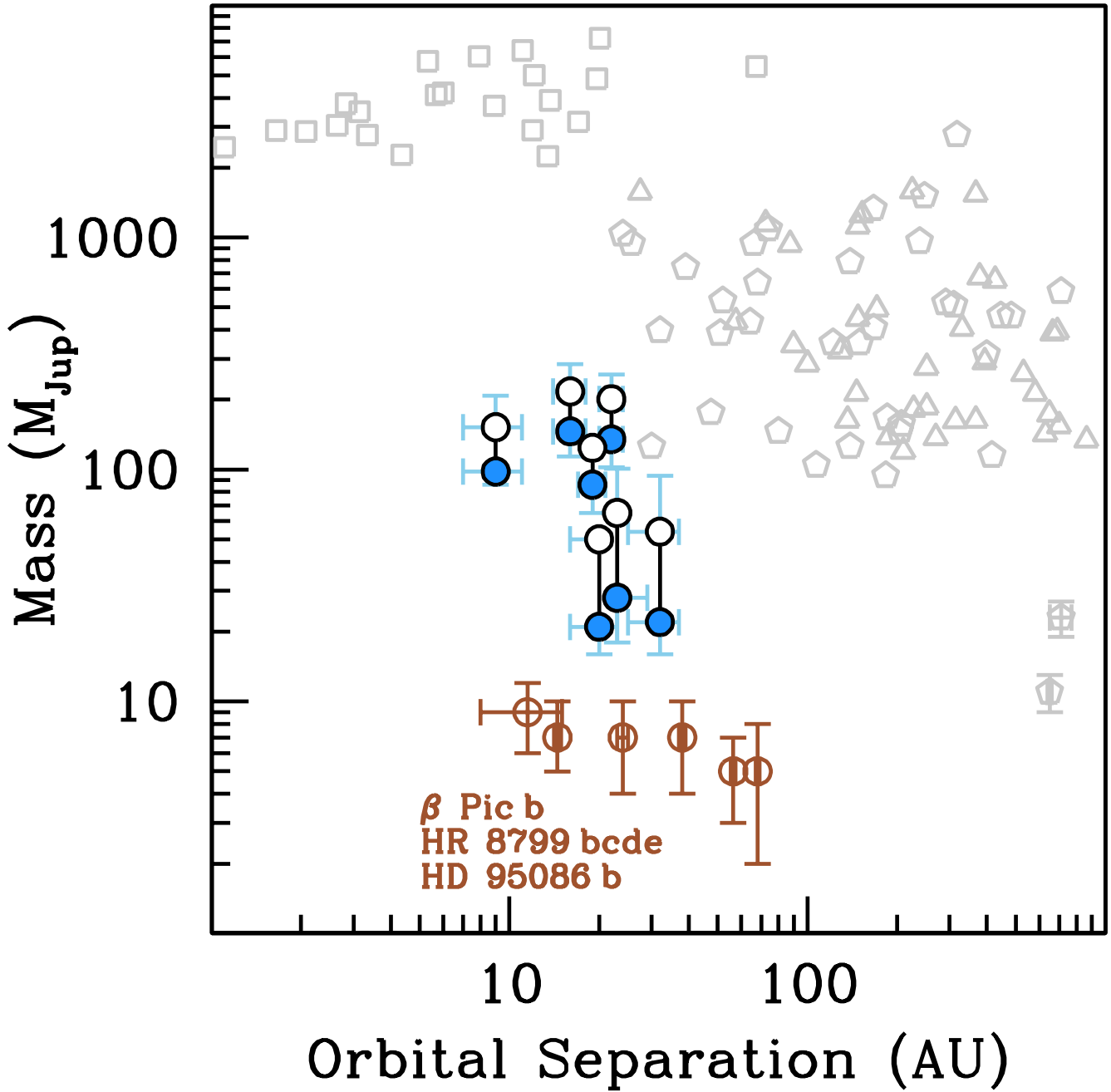}
  \caption{Companions to Sco-Cen stars of spectral type B, A, and F  ($\gtrsim$1.5$M_{\odot}$) expressed in terms of Jupiter masses (\mjup) and projected orbital separation (AU).  The blue and white circles denote our new detections of companions to Sco-Cen BAF stars using aperture masking interferometry (see Table\,\ref{contrasttable}), assuming a lower limit age of 10 Myr (blue circular points), and an upper limit age of 20 Myr (white circular points).   The gray symbols denote previous identifications from the literature of companions to Sco-Cen BAF stars obtained through conventional AO imaging (triangles, Kouwenhoven et al.~2005), interferometry (squares, Rizzuto et al.~2013) and more recent imaging studies  \citep[pentagons,][]{jlj13, ljv14, bmr14}.   For comparison, the brown circular symbols indicate the planetary mass companions to the young A-stars HR\,8799, $\beta$\,Pic, and HD\,95086, respectively \citep{mzk10, lbc10, rameau95086}}
  \label{parspace}
\end{figure*}

\section{Results}\label{results}
In Table\,\ref{contrasttable} we present the key properties of our newly detected companions, including the relative $L'$-band contrast ratios, angular separations, position angles, absolute magnitudes for the detected companions, as well as confidence levels of each of the detected companions to these Sco-Cen BAF stars.   For masses less than 0.1\,\msun, we convert these absolute magnitudes into physical masses for the DUSTY models \citep{cba00} corresponding to ages of 10 and 20 Myr.   For masses greater than 0.1\,\msun, we interpolate the \citet{bca98} models, assuming a mixing length parameter $L_{mix}$=1.9$H_p$, with $H_p$ denoting the pressure scale height. Our $L'$-band photometry for the host stars, used to calculate the relative brightness of the companions, is obtained from the WISE W1 channel. 

Figure\,\ref{parspace} shows our newly detected objects in a mass versus semi-major axis diagram, where we plot the lower (upper) limit age of 10 Myr (20 Myr) for each object using the values tabuated in Table\,\ref{contrasttable}. For context, Figure\,\ref{parspace} also shows several other detections of objects orbiting Sco-Cen stars of spectral type B, A, and F using conventional AO \citep{kbz05, jlj13, ljv14, bmr14}.     Also in the figure, we also show recent  interferometric detections of companions with primarily stellar masses ($\gtrsim$1\,\msun) taken from \citet{rir13}. In addition, the Figure shows the directly imaged exoplanets orbiting the closer, young ($\sim$10-30 Myr) A-stars HR\,8799, $\beta$\,Pic, and HD\,95086b \citep{mzk10,lbc10,rameau95086}.  Our objects occupy a similar orbital range (10 to 30 AU) to these directly imaged exoplanets, but with larger masses. Some works \citep[e.g.][]{vpm12} have suggested the peak of the companion distribution lies in this orbital range.

Three of our newly detected objects, HIP\,73990B, HIP\,73990C, and HIP\,74865B, have mass ratios clearly below the Hydrogen burning limit ($\sim$72$\,M_{Jup}$), even assuming the older 20\,Myr age.   These are objects at 10--30\,AU that unambiguously occupy the so-called ``brown dwarf desert'' \citep[e.g.][]{kbk07, kim08,kim11}, an observationally determined dearth of brown dwarf objects traditionally categorized as having $q$\,$\lesssim$\,0.1. Any detection of objects in this mass range will be particularly important to inform theoretical and numerical models of multiple systems \citep[e.g.][and references therein]{b09}.  

Given the very small angular separations at which aperture masking interferometry performs ($\lesssim$0.25$^{\prime\prime}$), the likelihood of contamination from background stars is negligible. For our survey, the number of expected contaminants has been estimated based on the local surface density of stars in the vicinity of the targets in our sample. This density was estimated using both the 2MASS survey and a star count algorithm that combines a spatial and luminosity model for the Milky Way (for the thin/thick disk, halo, bulge and present-day mass function \citep{rgh02}.   Nonetheless, to obtain the number of expected false detections in our broad survey of 140 stars, we extrapolate the false alarm rate from \citet{kim08}, which predicts 0.3 false detections for 60 Sco-Cen stars.  With this rate in hand, for our survey of 140 stars, we would expect $\sim$0.7 false detections.  Thus, the probability of identifying seven companions, as we have done in this work, would require a false alarm rate an order-of-magnitude greater than that of \citet{kim08}. Such a circumstance is exceedingly unlikely, since the current study also focusses on the Sco-Cen region.  Thus, the conventional need to confirm common proper motion is much less urgent.  Nonetheless, we will continue to monitor these targets with the goal of fully characterizing the orbital motion of the companions, as well as establishing common proper motion with the host stars. 

Of the six stars targetted in this study, only HIP 73990 has evidence for significant excess emission at 22\,$\mu$m as measured by the NASA WISE mission ([3.6]-[22]\,$\mu$m = 1.62), suggesting the presence of a debris disk. At the same time, none of our six target stars have any statistically significant excess emission at 4.6\,$\mu$m or 12\,$\mu$m, which would suggest the presense of more optically thick, protoplanetary disks. HIP73990 is also our only target with more than one detected companion. We devoted several additional post-processing tests to ensure the dual detections were not spurious (e.g. optical ``ghosts'' in the field of view, a companion present around another mutual calibrator star, etc.).  A thorough reprocessing of the data while varying the calibration scheme revealed that both detections were present irrespective of the number of mutual calibrators stars used.  Furthermore, the NACO L27 camera and $L'$-band filter have well characterized optical ghosts present in the focal plane. However, the position of these artifacts are fixed, and are easily excluded by a mask. Nonetheless, as an additional check, we performed a data reduction using only those files with the interferograms in identical places on the focal plane. The detection of the companions were robust against these changes as well.  Lastly, we rule out the possibility that the closure phase signal could perhaps be caused by aliasing from a more distant source, since \citet{jlj13} do not mention any detection of faint companions in their observations of this star using conventional AO imaging.  

\section{Conclusions \& Summary }\label{conclusions}
In this work, we report the detection of seven companions to intermediate-mass stars of spectral types B, A, and F in the Sco-Cen association.  With assumed ages of 10-20 Myr, our newly detected objects are observed shortly after the epoch of formation, residing on orbital scales comparable to objects in our solar system ($\sim$10-30\,AU). The young age of these stars allows detection of brown dwarf (and potentially planetary) mass objects even with modest achieved contrasts $\Delta L'$$\approx$4-6 mags. We highlight our main findings in this work as follows: 

\medskip

1) All of our newly detected companions are unambiguously ``brown dwarf desert'' objects with mass ratios $q$$\sim$0.01-0.08.  The derived masses of our newly detected companions suggest they are more massive analogs of the planetary mass companions HR8799\,bcde,  \citep{mzk10}.  HIP 73990 is perhaps the strongest analog for HR 8799, with multiple detected objects with masses $\sim$20-50 $M_{\rm Jup}$ ($q$\,$\approx$\,1--3\%).  Along with HIP 74865B, these three objects are all clearly below the Hydrogen burning limit, even assuming an age of 20 Myr.   

\medskip

2) The objects presented in this paper have much smaller orbital separations than the previously reported substellar companions to Sco-Cen BAF stars detected through conventional AO \citep{kbp07,jlj13, ljv14}.  

\medskip

3) The infrared brightnesses presented here will further serve as luminosity measurements of very young objects against which evolutionary models can be compared, thereby constraining the initial entropy of forming low mass companions \citep[e.g.][]{mc14}. 

\medskip

4) This study also begins to fill an important gap in our knowledge of the multiplicity of intermediate mass stars at young ages \citep[e.g.][]{dcb04}.  Specifically, some studies suggest \citep[e.g.][]{kbp07} that multiplicity is an {\it essential} outcome of the formation of intermediate mass stars, serving as a reservoir for the conserved initial angular momentum. 

\medskip

5) The new companions presented in this paper will be prime targets not only for follow-up spectroscopy \citep[e.g.][]{hbv15}, but also to search for fainter companions at large separations using the latest generation of dedicated exoplanet imagers such as GPI and SPHERE. Finally, as GAIA parallaxes are derived for these objects, follow-up high resolution host star spectroscopy will be highly beneficial to better determine the host star physical properties, such as \teff, $\log$$(g)$, and metallicity.   

\acknowledgments
We thank the anonymous referee for several helpful comments. We also thank Isabelle Baraffe and Gilles Chabrier for producing versions of their evolutionary models at customized ages. This work was performed in part under contract with the California Institute of Technology (Caltech) funded by NASA through the Sagan Fellowship Program and an NSF Astronomy and Astrophysics Postdoctoral Fellowship under award AST-1203023.  ALK was suported by a Clay Fellowship as well as NASA through Hubble Fellowship grant 51257.01 awarded by the STScI, which is operated by AURA, Inc., for NASA, under contract NAS 5-26555.  Some of the data presented herein were obtained at the W.M. Keck Observatory, which is operated as a scientific partnership among the California Institute of Technology, the University of California and NASA. The Observatory was made possible by the generous financial support of the W.M. Keck Foundation.

\end{document}